\documentclass[12pt]{article}

\usepackage{graphicx}

\newcommand {\beq}{\begin{equation}}
\newcommand {\eeq}{\end{equation}} 
\newcommand {\beqr}{\begin{eqnarray}}
\newcommand {\eeqr}{\end{eqnarray}} 
\newcommand{\rf}[1]{(\ref{#1})}

\newcommand{\der}[2]{{\fr{ \textstyle \partial #1}{ \textstyle \partial #2}}}

\newcommand{\fr}[2]{{\frac{ \textstyle #1}{ \textstyle #2}}}

\newcommand {\tm}{\times}

\newcommand{\rfg}[1]{ Fig.\ref{#1}}

\usepackage{pdftricks}
\usepackage{amsmath}
\usepackage{wasysym}

\renewcommand {\deg}{^\circ}
\newcount\indi
\newcount\num

\begin{document}

\title{Variational Data Assimilation for Optimizing
Boundary Conditions in Ocean Models  }
\author{ Christine Kazantsev\thanks{
 Universite Grenoble Alpes, Laboratoire Jean Kuntzmann,  Grenoble, France },
 Eugene Kazantsev\thanks{INRIA, Laboratoire Jean Kuntzmann, Grenoble, France   },
Mikhail A. Tolstykh\thanks{Institute of Numerical Mathematics, Russian Academy of Sciences and 
Hydrometeorological Research Center of the Russian Federation, Moscow, Russia}
}
\date{}

\maketitle

\begin{abstract}
The review describes the development of ideas Gury Ivanovich Marchuk in the field of variational data assimilation for ocean models applied in particular in coupled models for long-range weather forecasts. Particular attention is paid to the optimization of boundary conditions on rigid boundaries. As idealized and realistic model configurations are considered. It is shown that the optimization allows us to determine the most sensitive model operators and bring the model solution closer to the assimilated data. 
\end{abstract}


{\bf DOI:} 10.3103/S1068373915060047

\section{Introduction}

We met G.I. Marchuk in 1990; at that time one of us started working in the Institute of Numerical Mathematics of Academy of Sciences of the USSR, another one was a postgraduate student in that institute, and
the third one was on probation in the institute after the thesis defense. By that time G.I. Marchuk had left the
post of the President of Academy of Sciences of the USSR and had an opportunity to pay more attention to
the activities of the institute. We were amazed how quickly he could gain an understanding of a paper or a
report even if they touched a subject being new for him.

G.I. Marchuk freely shared his experience of a researcher and human. For example, one of his recommendations was to change the research subject every few years. He made this himself: the scope of his scientific interest ranged from the modeling of nuclear reactors to immunology. The results of his studies in
each area formed a book. His monographs included only new information that had never been published
before. Each monograph is a book of new ideas, explanations, and computations.

One of the lectures by G.I. Marchuk in Institute Elie Cartan in Nancy dealt with the theory of optimum
control, adjoint operators, and theory of perturbations. He worked a lot on the methods of the solution of inverse problems, especially of complex ones for nonlinear systems of equations. In the middle of the 1970s
in [1, 44] he formulated the theory of perturbations for solving such problems and 20 years later [2, 4] told
how this theory was developed by him and his followers.

The followers of G.I. Marchuk used the theory of perturbations for initializing atmospheric models by
solving the problem of optimum combination of the data obtained from the model and from observations
[5, 6, 45]. Soon after that, based on the papers by G.I. Marchuk and his friend J.-L. Lions on the optimum
control [37], French scientists also used adjoint equations for working out the variational algorithms of assimilation of meteorological data and for optimizing the initial conditions of atmospheric models [31, 33].

By the late 1980s--early 1990s, many research teams including both mathematicians and geophysicists
were involved into the development of new methods of data assimilation and solution of inverse problems
[3, 18, 38, 46, 47] as well as into the use of these methods in the models of the ocean and atmosphere [13,36, 49, 56]. This resulted in considerable increase in the reliability of the weather forecast [23]. In the second half of the 1990s, the European Center for Medium-Range Forecasting (ECMWF), Met Office, Meteo
France, and many other centers started the operational use of the practical implementation of the theory and
methods of data assimilation.

The models of the atmosphere with the fixed sea surface temperature and sea ice concentration have
been commonly used for the short- and medium-range numerical weather prediction. In the world the probabilistic long-range weather forecasting mainly use the coupled models of the atmosphere, ocean, sea ice,
and active soil layer [15]. For example, in ECMWF and Met Office, the NEMO ocean model is used jointly
with the corresponding models of the atmosphere (http://www.nemo-ocean.eu/) [54]. This model is described in detail in [43] and is widely used by the European community, in particular, in the coupled models
intended for the long-range weather forecasting.
Recent studies have demonstrated that the use of coupled models is essential for the medium-range and
even short-range forecasting at high latitudes [50]. This is also associated with the role of polynyas in the
formation of mesoscale polar cyclones. Currently the leading centers, ECMWF and Met Office, are planning to carry out the operational implementation of the coupled model for the medium-range forecasting.

Due to the limited number of observations in the ocean, the ocean models were traditionally tested only
for the simulation of certain integral characteristics. The use of the ocean–atmosphere coupled models for
the weather forecasting dictates the heightened (as compared to the climate models) requirements to the accuracy of the simulation of the fields of sea surface temperature and sea ice concentration. Therefore, the
development of the system of observational data assimilation for the optimization of conditions on rigid
boundaries for the ocean models considered in the present paper is also an important issue for the systems
of numerical weather prediction, especially because the coupled ocean–atmosphere model intended for the
long-range weather forecasting is developed in Russia [7].

It should be noted that the assimilation of observational data was first of all aimed at the optimization of
the initial conditions of models. As demonstrated in [39], the solution of the nonlinear model is extremely
sensitive to initial data and the small error in their estimate may increase exponentially. This fact attracted
attention to studying the resistance of the solutions of geophysical models to inaccuracies in initial conditions and defined the follwing main objective of data assimilation: the three- and four-dimensional analysis
of the fields of meteorological and oceanic data in order to agree them with the model and use subsequently
as initial data for the model.

Certainly, it is doubtless that the solution of the nonlinear model is highly sensitive to initial conditions.
However, one can suppose the same or even higher sensitivity of the solution to the other parameters of the
model such as the diffusion coefficient, underlying surface approximation, the parameterization of subgrid
processes, the description of near-continent areas in the ocean models, and other internal parameters of the
model. In fact, the strong dependence of model currents on boundary conditions was noted in [8, 57], on
bottom topography approximation, in [17, 22, 41], on diffusion coefficients, in [11], and on fundamental
parameterizations such as the Boussinesq approximation and hydrostatics, in [40]. The sensitivity of the
shallow water model to different parameters is analyzed in [30]. The theoretical and practical features of the
use of adjoint equations for assessing the parameters of geophysical models are discussed in [48]; however,
the comparatively small number of papers deal with data assimilation for the purpose of specifying or
optimizing the model parameters that differ from initial conditions. Several attempts can be cited of the use
of variational methods for optimizing bottom topography [25, 42] and for controlling the open boundary
conditions of regional ocean models [52, 53, 55]; however, the number of such papers is much smaller than
that of the papers dealing with the optimization of the model’s initial conditions.

\section{ Assimilating data to bring together a model and its boundary conditions. }

Let us dwell in more detail on the variational data assimilation in order to coordinate the model with its
boundary conditions on rigid boundaries. Special attention is paid to the situations with pronounced boundary layers because in this case the strongest dependence of the model solution on the approximation of
boundary conditions is observed.

The optimization of boundary conditions using adjoint equations has already been used for simple systems. For example, the optimum control of conditions is described in [12, 19] for the linear heat equation, in
[34, 35], for the Burgers equation, and in [32], for the nonlinear balance equation. In [26] it was proposed to
control the discretization at the boundary points of the differential operator that uses them. The proposal
followed from the result [35] demonstrating that the approximation of nontrivial boundary conditions
should follow several rules in order to provide the needed order of approximation and the remainder. Indeed, the information about boundary conditions comes to the model through the specific discretization of
operators near the boundary. Control over this discretization allows one, on the one hand, to control the
conditions and, on the other hand, to control their approximation to the model grid.

Let us suppose that the derivative operator at the staggered grid for the function and its derivative is defined at $x > 0$ and the certain boundary condition is formulated at the point $x = 0$. Then, if the function is defined at the integer points $x_n=nh$  ($n$ is an integer number), let us use the following formula for computing
the value of the derivative near the boundary:

\beq
\biggl(\der{u}{x}\biggr)_{x=h/2}= \fr{ \alpha_0+\alpha_1 u(0)+\alpha_2 u(h) }{h}
\label{s0}
\eeq
If the function is defined at the semi-integer points $x_n=(n-1/2)h$, the derivative is computed using the
\beq
\biggl(\der{u}{x}\biggr)_{x=0}= \fr{ \alpha_0+\alpha_1 u(h/2)+\alpha_2 u(3h/2) }{h}
\label{s1}
\eeq

As clear, any boundary conditions including the inhomogeneous ones can be taken into account by
selecting the corresponding set of coefficients  $\alpha$. For example, the condition $u(0)=A$ will be taken into
account and approximated with the second order of accuracy by selecting $\alpha_0=-A,\; \alpha_1=0,\; \alpha_2=1$ in
formula  \rf{s0} and  $\alpha_0=-2A,\; \alpha_1=2,\; \alpha_2=0$ in formula \rf{s1}.

In view of the fact that the grid with staggered points is used, the problem solution requires interpolation
from integer to semi-integer points and vice versa. Interpolations at the boundary points should also take
account of boundary conditions; therefore, their discretization is carried out in a similar way by constructing the linear combinations of function values with coefficients $\alpha$ (that certainly differ from those used for
the discretization of derivatives).

Thus, the discrete approximation of derivatives and interpolation at the boundary points 
is presented in
the form of the linear combinations of the function values with coefficients $\alpha$. On varying these coefficients
in the process of data assimilation, the cost function is minimized and the optimized values of these coefficients are found.

All coefficients  $\alpha$ are considered as independent variables. The specific set is used for each model parameter and each operator even if boundary conditions are identical because the approximation of identical
conditions to the grid may differ. Moreover, different sets of  $\alpha$ are used at the different points of the
boundary of two- or three-dimensional domain and the additional degrees of freedom are given for the
spatial approximation of the boundary.

Papers [26, 29] analyzed the optimization of coefficients  $\alpha$ in the process of the assimilation of the exact
solutions of the one-dimensional wave equation and linear shallow water model; the following conclusions
were made.

\begin{itemize}
\item The data assimilation used for controlling the operator discretization at the boundary points enables
one to compensate the errors of the numerical scheme and bring the numerical solution much closer to the
data having decreased the cost function by three or four orders; 
\item The adjoint equation and adjoint model turn out to be by 2–3 times longer than the adjoint model for
optimizing the initial conditions. This is manifested in the number of the code lines and in the computer
time needed for its computation and determines the heightened interest to the use of the procedures of the
automatic differentiation of the initial model code. It should be noted that at the differentiation of the complete models of the atmosphere or ocean, the obtained code of the adjoint model is to be optimized, especially in order to reduce its requirements to the main storage. An example of such optimization using the
TAPENADE differentiator [21] is given in [28]; there the storage requirements were lowered by 25 times.
\item  The control of operator discretization may lead to the non-uniqueness of the result and to the existence of the nonzero core and does not allow its use for identifying boundary conditions as parameters in
the general case.
\item Such control may result in the failure of physical hypotheses made in the process of the model construction. For example, in the case of the one-dimensional wave equation, the initial geometry of domains
in the shallow water model was changed not quite soundly in order to compensate the errors of the numerical scheme inside the domain.
\end{itemize}

Thus, the control of operator discretization at the boundary points can indicate a simple way of bringing
the model solution closer to the data. However, such control needs the additional analysis of the results
from the point of view of uniqueness and physical acceptability of obtained approximations.

In [27, 28], the optimization of coefficients a was carried out for the NEMO (Nucleus for European
Modelling of the Ocean) European ocean model [43]. The core of this model is equations for two
components of velocity, for temperature and salinity, and for sea surface height. The use of the hydrostatic
approximation causes the fact that the vertical velocity is a diagnostic variable.

Let us start the experiments with the NEMO model with the simplified configuration: the rectangular
parallelepiped with the size of $30\deg$  along the longitude, $20\deg$ along the latitude, and 4195 m along the vertical
(depth). The horizontal resolution of the grid is  $0.25\deg$. As to the vertical, all the depth is divided into four
layers of equal thickness. Thus, the grid consists of $120\tm 80\tm 4$  points for every model variable. The model
is discretized at the three-dimensional analog of the Arakawa C-grid [9].

The wind friction stress is applied to the sea surface and is defined by the formula:
$$
\tau_\phi=0, \quad \tau_\lambda=-0.1\fr{N}{m^2}\tm \fr{\cos(\pi(\phi-24\deg))}{44\deg-24\deg}, \label{forc}
$$
The conditions of impermeability and slipping are used on the lateral boundaries:
$$
\vec{U}\cdot\vec n=0,\quad \der{\vec{U}\cdot\vec{\tau}}{\vec{n}}=0,
$$
(where $\vec{U}=(u,v)$ is the vector of horizontal velocity and $\vec n, \vec{\tau} $ are the normal and tangent to the domain
boundary.

This academic configuration is of special interest for studying the effects of boundary conditions and
their approximation on the model solution in the presence of strongly pronounced boundary layers. The
non-triviality of the approximation of boundary conditions is clearly manifested in the experiment when the
grid is not parallel to the parallelepiped walls. In this case, especially in the presence of boundary layers, the
whole solution turns out to be strongly depending on how the domain boundaries are approximated by the
grid. This problem has been actively investigated in recent 15 years (for example, [8, 16, 20]); however, no
recommendation have been made so far on the approximation of such boundaries for finding the reasonable
solution in the general case.

This problem is rather important in oceanography because the continental margins most often do not coincide with the grid of finite-difference models that hampers their correct approximation, especially in the
presence of boundary layers and jet streams such as the Gulfstream and Kuroshio. Certainly, this problem
can be solved by rejecting finite differences and by using the methods that are more adapted to the complex
geometry (for example, finite elements [14] or finite volumes [51]). However, the use of finite differences
currently prevails due to their simplicity and speed.

Let us gain from the simplicity of the described rectangular configuration for the experiments with the
optimum control of operator discretization at the boundary points. If the grid is parallel to the lateral boundaries, these boundaries turn to be approximated by straight lines. If the grid is rotated by $45^\circ$, the lateral
boundaries are approximated by a “staircase” that results in the not quite correct formulation of boundary
conditions, in particular, of slipping conditions. We know that the physical boundary is a straight line.
Hence, according to the slipping conditions, the nonzero tangential velocity can be observed on the boundary because only the normal component must be zeroed due to the impermeability condition. However, the
“staircase” approximation results in the fact that the impermeability condition is applied to the velocity.
The slipping is taken into account only by zeroing the relative vorticity at the boundary points.

After the model have been integrated during 5 years at the grid being parallel to the lateral boundaries
until the statistically stable state is reached, the field of currents is obtained with the strongly pronounced
boundary layer close to the western and northern boundaries and with the poorly pronounced wave activity.
Let us consider the obtained solution standard. The sea surface height for this solution is demonstrated in \rfg{sg-ref}A. 

Now let us rotate the grid by $45^\circ$  and again integrate the model for 5 years with the same diffusion, forcing, and boundary conditions strictly following the approximations embedded into the NEMO ocean
model. The topography of the surface of this solution is presented in Fig. 1B. This solution differs from the
standard one. Due to the “staircase” approximation of the boundary, the boundary current in the north is reduced by more than twice that has already been noted several times [8, 16].

\begin{figure}[h]
  \begin{center}
  \begin{minipage}[l]{0.48\textwidth}
   A. \\ 
  \centerline{\includegraphics[angle=0,width=0.99\textwidth]{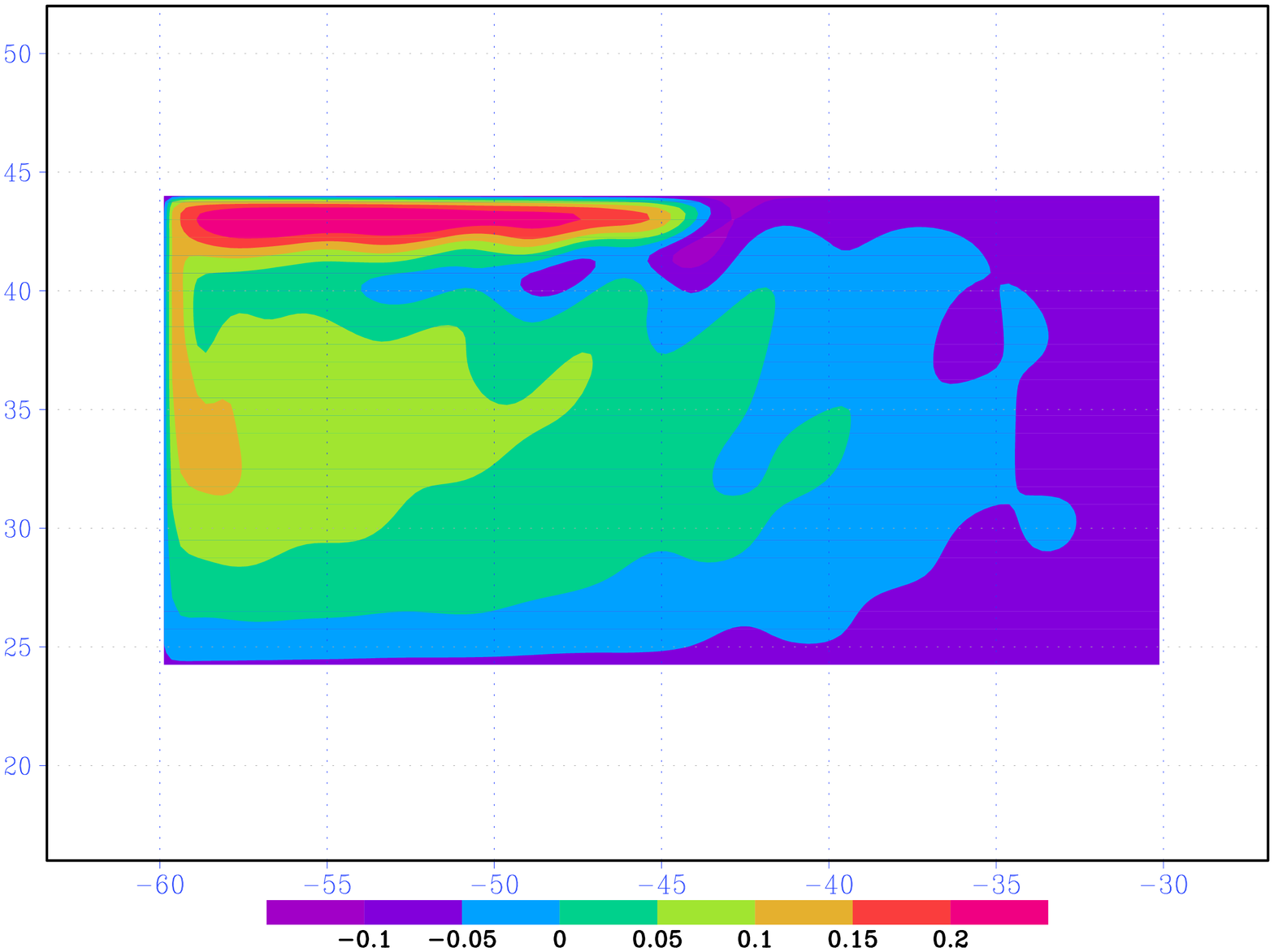}}
  \end{minipage} 
  \begin{minipage}[r]{0.48\textwidth} 
      \hfill B. 
  \centerline{\includegraphics[angle=0,width=0.99\textwidth]{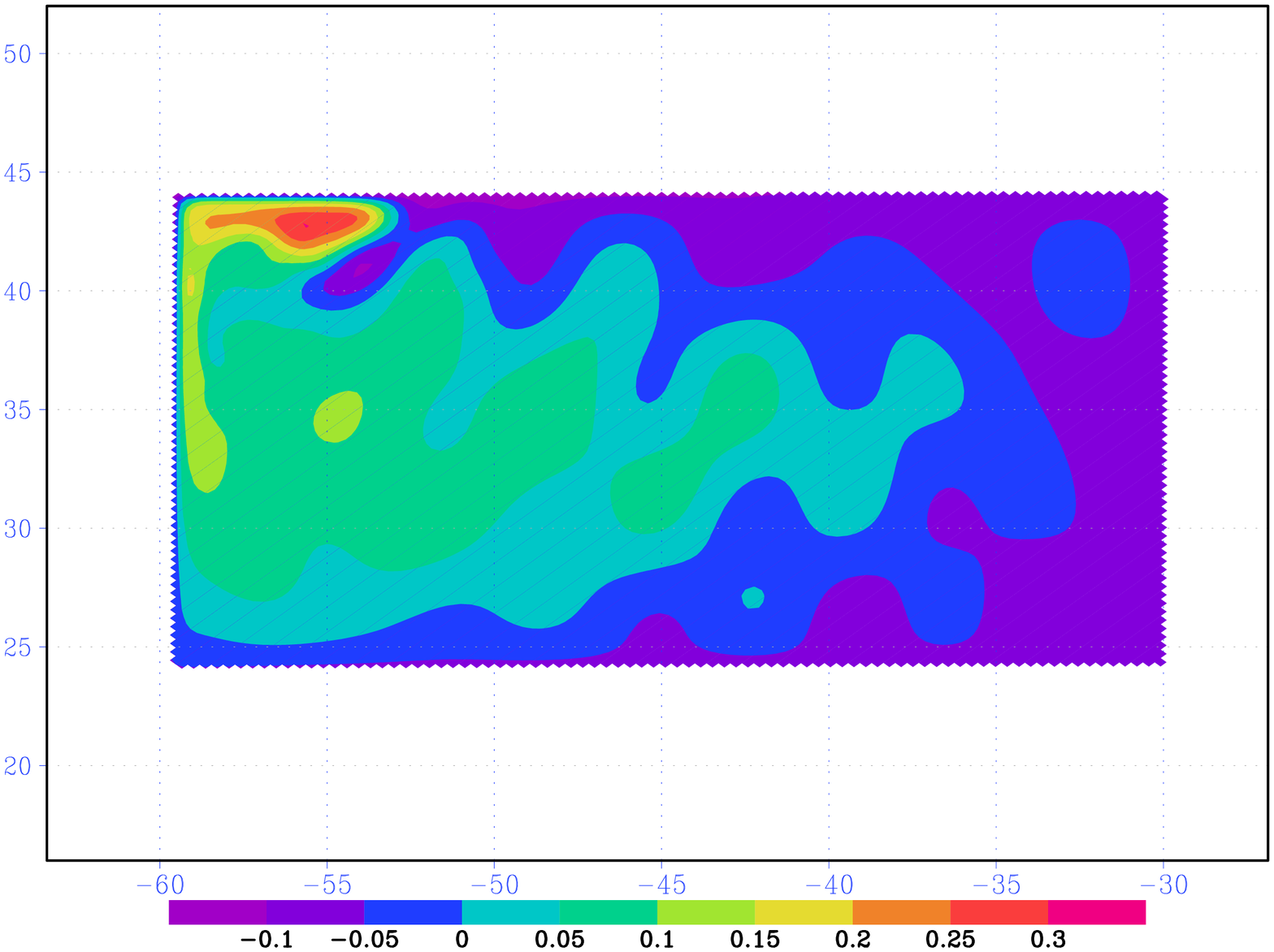}}
  \end{minipage}  
  \end{center} 
  \begin{center}
  C.\\
  \begin{minipage}[l]{0.48\textwidth}
 \centerline{\includegraphics[angle=0,width=0.99\textwidth]{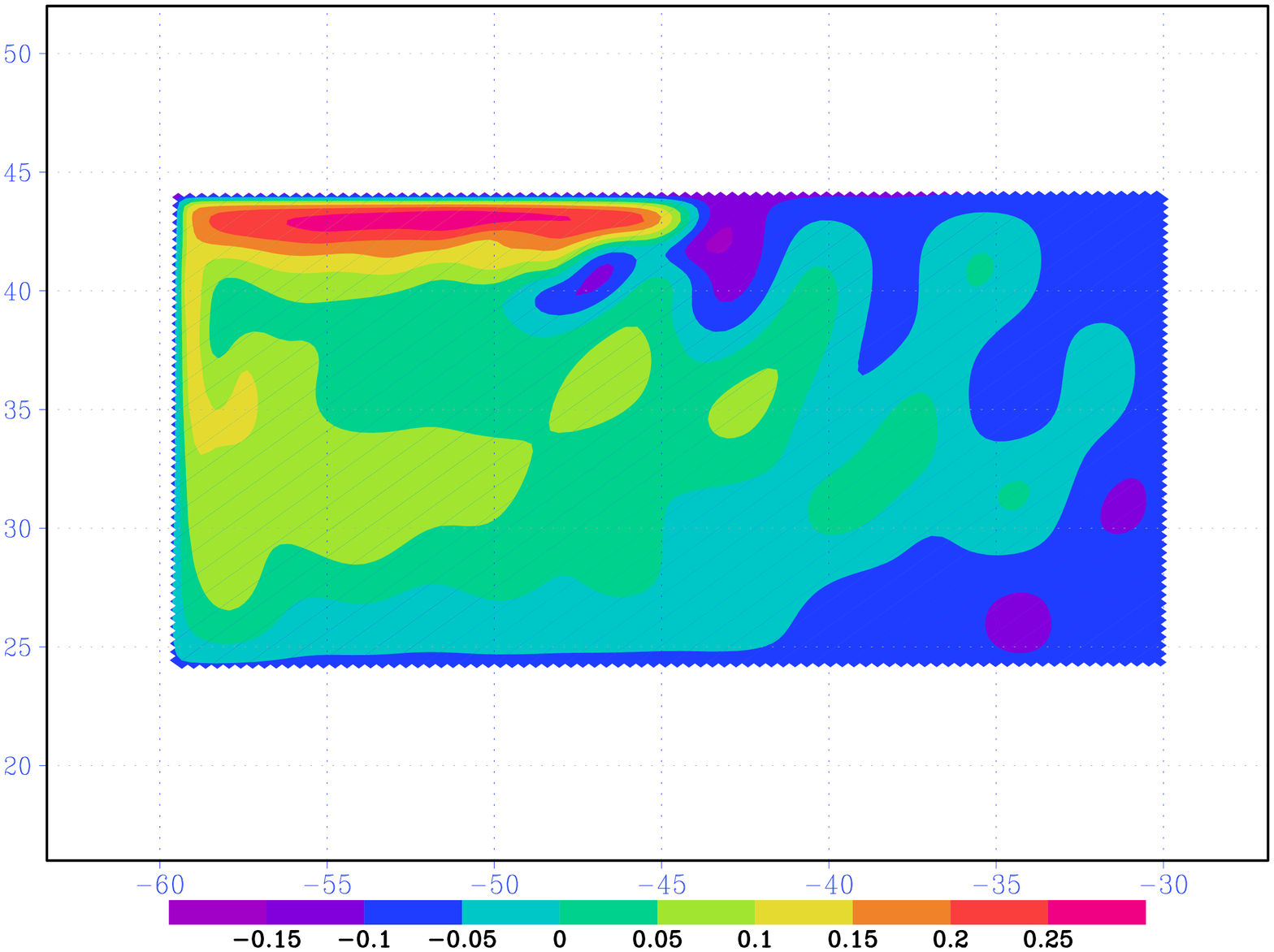}}
   \end{minipage} 
  \end{center} 
  \caption{The sea surface height in the solution of model equations  at the parallel grid (A), at the grid rotated by $45^\circ$ (B), and at
the rotated grid with the optimized discretization  (C). }
    \label{sg-ref}
\end{figure}

Let us assimilate the standard solution interpolated to the rotated grid in the experiment on data assimilation and operator discretization control at the boundary points. The assimilation window $T=50$ days
turns out to be sufficient to provide the stability of obtained discretizations. In all derivatives and interpolations of the model for three coordinates at all boundary points, let us replace the classic discretization formulae by  (1) and (2) containing the controlled coefficients  $\alpha$. The series of experiments on the
standard solution assimilation revealed that the model solution at the rotated grid is sensitive only to the
discretization of derivatives in the vorticity operator. All other interpolations, the first and second
derivatives of the model did not practically change the initial coefficients  $\alpha$ whereas the computation of
vorticity at the boundary points was carried out using other formulae: instead of the usual approximations
of derivatives with  $\alpha= \pm 1$, we get the values $\alpha= \pm 1.4, \pm 0.6$ at the nearest point to the boundary. For
example, the vorticity close to the western boundary is computed as

$$
\biggl(\der{v}{x}-\der{u}{y}\biggr)_{i,j}= \fr{ 0.6 v_{i,j-1} -1.4 v_{i-1,j-1}}{h}-\fr{ 1.4 u_{i,j} -0.6 u_{i,j-1}}{h}.
$$
In other words, the optimization requires the supplementation of the usual approximation of derivatives $\delta_x\left[v\right]= \fr{ v_{i} - v_{i-1}}{h}$ with the projection of velocity to the tangent to the boundary divided by $R$
\beq
\biggl(\der{v}{x}-\der{u}{y}\biggr)_{i,j}=\delta_x\left[v\right]-\delta_y\left[u\right] - \fr{0.8}{h}(\overline{u}+\overline{v})=\delta_x\left[v\right]-\delta_y\left[u\right]+\fr{ \vec{U}\cdot\vec{\tau}}{R},
\label{modf}
\eeq
where $R=-\fr{h}{0.8 \sqrt{2}}$. 

To get the meaning of this additive, let us advert to papers [10, 57] that demonstrate that the formulation
of boundary slipping conditions on the curvilinear boundary requires such additive. The parameter $R$ has a
meaning of the local radius of curvature of the boundary.

Having discretized the derivatives in the equation for vorticity at the boundary points following (3), we
get the solution for the height of the sea surface whose topography is demonstrated in Fig. 1C. As clear, it is
quite similar to the sea surface height of the standard solution in Fig. 1A; this enables a conclusion that the
effects of “staircase” approximation of the boundary are compensated. The optimized boundary turns out to
be not a straight line as it could be expected but a certain curve with the constant negative radius of curvature close to $-h/ \sqrt{2}$. Moreover, as demonstrated in [27], in the case of the inclined grid the formulation of
slipping conditions on the straight physical boundary of the domain may result in the instability in the approximation of the Coriolis parameter.

Data assimilation for the more realistic situation was carried out in [28].This paper uses the ORCA-2
configuration of the NEMO model that was worked out for simulating the World Ocean at the grid with the
horizontal resolution of  $2^\circ$  and with 31 vertical layers. In spite of the low resolution, this configuration is
widely used for studying the ocean dynamics.

In this experiment the observational data from ECMWF databases received from Jason-1 and Envisat
satellites were assimilated. Besides, the profiles of temperature and salinity from the ENACT/ENSEMBLES data bank were assimilated as well. The measurements of the sea surface height contain about
112000 values obtained on January 1–20, 2006. The profiles of temperature and salinity are represented by
the data of 200000 measurements during the same period.

As well as in the previous experiment, let us first of all determine the operators whose discretization
affects the model solution most of all. It is revealed that the discretization of zonal and meridional operators
affects the results almost in no way. In the experiment that allows controlling only the discretizations of
horizontal operators, the cost function does not virtually decrease (less than 1\%) and there is almost no defference between model solutions with the optimized and classic discretizations. This fact is easily explained
by the low horizontal resolution of the model. The two-degree grid of ORCA-2 configuration needs the
high horizontal dissipation whose coefficient is equal to  $40000\fr{\mbox{m}^2}{\mbox{s}}$ in the extratropical zone. Such significant smoothing quickly levels potential boundary effects. The modification of the boundary in the previous
experiment led to the change in the model solution under other conditions: the higher resolution and lower
horizontal dissipation ($200\fr{\mbox{m}^2}{\mbox{s}}$).

The maximum influence on the model solution is exerted by the boundary conditions in the operator of
vertical dissipation. This is true for the ocean bottom only. Their optimization modifies considerably the
fields of velocity, temperature, and salinity not only in the bottom layer but also on the sea surface that is
proved by the sea surface height in the North Atlantic (Fig. 2).

To obtain this result, we integrated the ORCA-2 configuration of the NEMO model from January 1 to
30, 2006 using the initial data and the external forcing obtained from the observational data for that period.
The sea surface height at the end of integration is presented in Fig. 2A. After that the assimilation of
altimetric data and profiles of temperature and salinity for 20 days (January 1 to 20) was carried out and the
optimized coefficients $\alpha$ were found which define the discretization of operators $\der{}{z}A^z\der{u}{z} $ и $\der{}{z}A^z\der{v}{z}$ at bottom points. Using these coefficients, the model was integrated for 30 days once again and the solution
was found corresponding to the sea surface height demonstrated in Fig. 2B.

\begin{figure}[h]
  \begin{center}
    \begin{minipage}[l]{0.48\textwidth}
   A. \\
   \vspace{-2mm}
  \centerline{\includegraphics[angle=-90,width=0.99\textwidth]{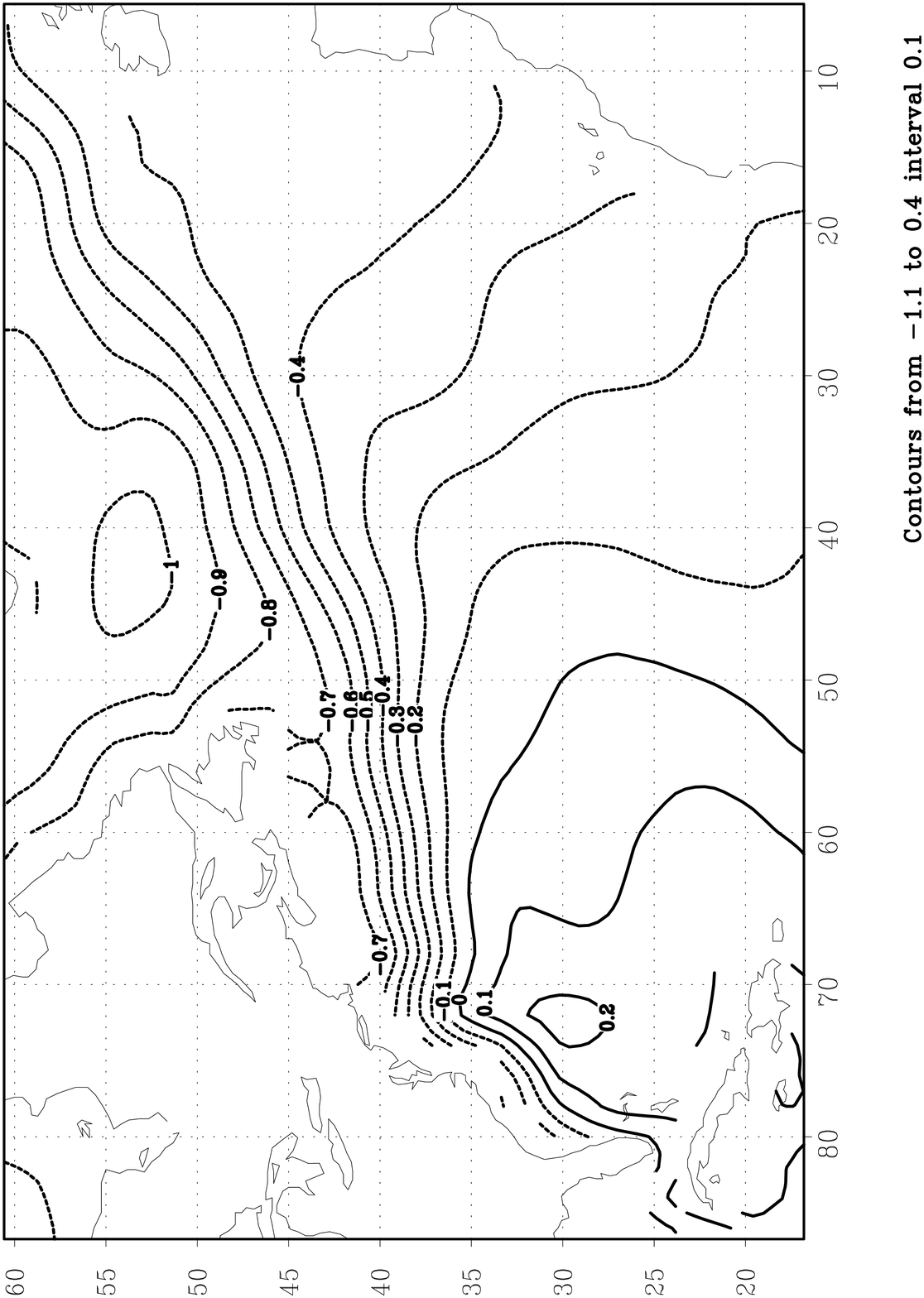}}
  \end{minipage} 
  \begin{minipage}[r]{0.48\textwidth} 
      \hfill B. 
  \centerline{\includegraphics[angle=-90,width=0.95\textwidth]{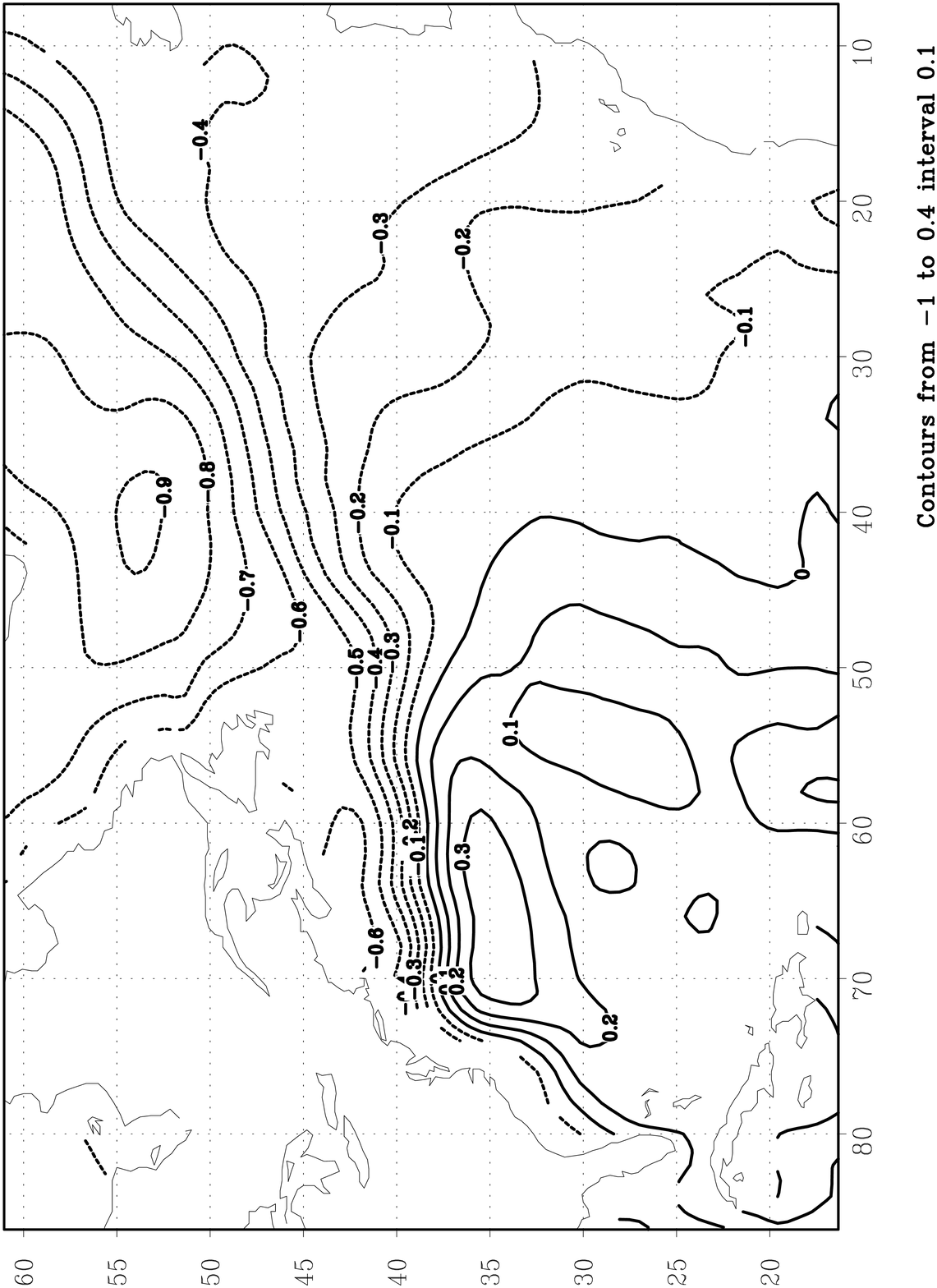}}
  \end{minipage} 

  \end{center}   
    \caption{  The sea surface height in the North Atlantic on January 30, 2006 computed using  the classical (A) and optimized (B) 
discretization of operators of vertical diffusion at bottom points.}
\label{opiz}
\end{figure}

The basic feature of the ocean circulation in the North Atlantic is a formation of a narrow and long jet
stream, the Gulf Stream. Certainly, the Gulf Stream is present in the satellite altimetry whose data are assimilated by the model. Therefore, the jet stream becomes more pronounced in the model solution after
data assimilation: the velocity (being proportional to the meridional gradient of sea surface height) and the
length of the Gulf Stream increase. However, it is difficult to explain why the modification of boundary
conditions for the zonal and meridional velocities in the certain areas of the bottom is the most efficient for
intensifying the surface jet stream.

\section{Conclusions}

Thus, it is clear that boundary conditions on rigid boundaries are also important in realistic ocean models. The use of the variational assimilation of data for their optimization can be useful in spite of the high
cost of this procedure. As mentioned, data assimilation defines the model operators being the most sensitive
to boundary conditions as well as identifies the geographic regions where such sensitivity is most pronounced. We automatically provide the stability of the obtained discretization by choosing the rather durable assimilation window. Its conservative properties can also be provided by adding the corresponding
member to the cost function as demonstrated in [24]. However, it is often needed to comprehend the obtained results from the point of view of the correspondence to the accepted physical imperatives and hypotheses. The understanding of the physical sense of optimized coefficients for realistic models can also be
hampered.

In conclusion it should be noted that the present review covers only the small part of the research areas
developed by the followers of G.I. Marchuk. During his outstanding scientific life, G.I. Marchuk laid a
strong foundation for studies in different fields of science and technology. Besides, he created a strong scientific school. Nowadays his grateful followers, “scientific grandchildren and great-grandchildren,” develop his ideas and continue the studies which he initiated in different fields of science, in different countries, and for different applications.

His work lives on and is continued in scientific papers, models, and technologies; it opens up new vistas
for the mankind. His memory is kept with care by all who had the luck to know him personally and to work
with him.

\end{document}